\documentclass[a4paper]{spie}  

\usepackage{graphicx}

\usepackage{xspace}
\usepackage{siunitx}
\usepackage{amssymb,amsmath,amsfonts}
\usepackage[colorlinks=true, allcolors=blue]{hyperref}
\usepackage{booktabs}
\usepackage{ulem}
\usepackage[T1]{fontenc}

\usepackage{upgreek}

\newcommand{\superbit}{\textsc{SuperBIT}\xspace}

\newcommand{\coord}[1]{{\mathcal{F}}_{#1}}

\title{The design and development of a high-resolution visible-to-near-UV telescope for balloon-borne astronomy: SuperBIT}

\author[a]{L. Javier Romualdez}
\author[c]{Steven J. Benton}
\author[d]{Paul Clark}
\author[a]{Christopher J. Damaren}
\author[e]{Tim Eifler}
\author[c]{Aurelien A. Fraisse}
\author[b]{Mathew N. Galloway}
\author[b]{John W. Hartley}
\author[c]{William C. Jones}
\author[a]{Lun Li}
\author[b]{Leeav Lipton}
\author[c]{Thuy Vy T. Luu}
\author[d]{Richard J. Massey}
\author[b,g]{C. Barth Netterfield}
\author[g]{Ivan Padilla}
\author[e,f]{Jason D. Rhodes}
\author[d]{J\"urgen Schmoll}

\affil[a]{University of Toronto Institute for Aerospace Studies (UTIAS), 4925 Dufferin Street, Toronto, ON, Canada}
\affil[b]{Department of Physics, University of Toronto, 60 St. George Street, Toronto, ON, Canada}
\affil[c]{Department of Physics, Princeton University, Washington Road, Princeton, NJ, USA}
\affil[d]{Centre for Advanced Instrumentation (CfAI), Durham University, Science Laboratories, South Road, Durham, UK}
\affil[e]{Jet Propulsion Laboratory (JPL), California Institute of Technology, 4800 Oak Grove Drive, Pasadena, CA, USA}
\affil[f]{California Institute of Technology, 1201 East California Blvd, Pasadena, CA, USA}
\affil[g]{Department of Astronomy, University of Toronto, 50 St. George Street, Toronto, ON, Canada}

\authorinfo{Further author information: (Send correspondence to L.J.Romualdez)\\ E-mail: javier.romualdez@mail.utoronto.ca, Telephone: 1-416-946-0946}

\begin{document} 
\maketitle

\begin{abstract}
Balloon-borne astronomy is unique in that it allows for a level of image stability, resolution, and optical backgrounds that are comparable to space-borne systems due to greatly reduced atmospheric interference, but at a fraction of the cost and over a significantly reduced development time-scale. Instruments operating within visible-to-near-UV bands ($300$ -- $900$ $\upmu$m) can achieve a theoretical diffraction limited resolution of $0.01''$ from the stratosphere ($35$ -- $40$ km altitude) without the need for extensive adaptive optical systems required by ground-based systems. The {\it Superpressure Balloon-borne Imaging Telescope} (``\superbit'') is a wide-field imager designed to achieve 0.02$''$ stability over a 0.5$^\circ$ field-of-view, for deep single exposures of up to 5 minutes. 
{\superbit} is thus well-suited for many astronomical observations, from solar or extrasolar planetary observations, to resolved stellar populations and distant galaxies (whether to study their morphology, evolution, or gravitational lensing by foreground mass). 
We report {\superbit}'s design and implementation, emphasizing its two-stage real-time stabilization: telescope stability to $1$ -- $2''$ at the telescope level (a goal surpassed during a test flight in September 2015) and image stability down to $0.02''$ via an actuated tip-tilt mirror in the optical path (to be tested during a flight in 2016). The project is progressing toward a fully operational, three month flight from New Zealand by 2018. 
\end{abstract}

\keywords{balloon-borne, wide field, high resolution, visible-to-near-UV, high gain control, stratospheric instrument}

\section{INTRODUCTION}
\label{sec:intro}

\subsection{Astronomical and Cosmological Background}

An inexpensive, near-space quality observing platform with instruments that can be replaced and easily upgraded has enormous potential for scientific discovery. While still in its early stages, NASA's recently developed super pressure balloon (SPB) capability can evolve into such a platform, especially in combination with high-resolution, wide-field imagers or spectrographs. The combination of diffraction limited angular resolution, extreme stability, space-like backgrounds, and long integrations enable transformative opportunities for astrophysics and cosmology. 

Numerous science objectives require these capabilities and have driven a highly oversubscribed demand for UV-visible-NIR observations on the Hubble Space Telescope (HST). While JWST (James Web Space Telescope) will provide a significant improvement over HST's capabilities longward of 600nm, and other projects, such as Euclid and WFIRST (Wide-Field Infrared Survey Telescope) will provide red and NIR (Near Infrared) imaging over wide areas, the inevitable demise of HST will leave us with no space-based capabilities in the blue and UV. Even in the visible and NIR, planned missions will not begin to exhaust the demand for wide-field, high-resolution imaging. SPB based imaging and spectroscopy can potentially advance a large fraction of research topics in astronomy and cosmology including: 

\begin{enumerate}
\item Exoplanets: searches and classification via imaging and spectroscopy.
\item Interstellar Physics: large-scale mapping of the interstellar medium; 
resolved imaging and spectroscopy of star forming regions; evolution of molecular abundances.
\item Stars: Young-star accretion variability; simultaneous star observations in the UV and NIR; astroseismology; search for pulsars in local group; search for white dwarfs in galactic halo.
\item Galaxies: Modified gravity constraints from galaxy morphology observations; galaxy evolution, e.g. a wide-field COSMOS survey; line-intensity mapping; study merger history and star formation triggers; study high-z galaxy morphology.
\item Black Holes: Low-mass AGN reverberation mapping in the UV-optical-NIR; variability of AGN; Gamma Ray Burst monitoring; black hole accretion disk dynamics; high-resolution imaging of blazars (highly compact, energetic radio sources).
\item Planetary Science: Comet water studies; UV comet imaging; wide-field survey for NEOs (Near Earth Objects); internal structure from Jupiter and Saturn via surface oscillations, study solar system moons, spectroscopy of planet atmospheres; deep imaging of Kuiper Belt Objects. 
\item Dark Energy/Dark Matter: Dark matter mapping of Active Galactic Nuclei (AGN); Strong Lensing time-delays; wide-field weak lensing to calibrate ground based shape measurements; combined imaging and spectroscopic weak lensing measurements.
\item Galaxy Clusters: galaxy cluster lensing mass surveys; supernovae observations; precision UV/visual photometric measurements complementary to future ground and space based missions.
\end{enumerate}

\noindent In the following we focus on the last two science topics. Dark Energy and Dark Matter together constitute 96\% of the content of the Universe's energy density, yet their nature remains poorly understood. 
Wide-field astronomical surveys during the next decade have potential to explore both concepts, via statistical measurements of e.g.\ large-scale structure, gravitational lensing, and redshift space distortions.
Amongst planned facilities, the ground-based Large Synoptic Survey Telescope (LSST) will explore a large volume of the Universe (18,000 deg$^2$ to 27.5 r-band magnitude), but will be limited by atmospheric seeing. 
Future space-based missions will have exquisite image quality but will be either substantially shallower or cover a smaller fraction of the sky. 
Euclid, which will overlap with LSST for $\sim$6000 deg$^2$, will be $\sim 2$ magnitudes shallower. 
WFIRST will be as deep as LSST, but its imaging will cover only 2300 deg$^2$. 

Balloon observations from flights at mid-latitude (45$^\circ$ S) provide an excellent opportunity for targeted research to build on these facility surveys, or for coordinated observations to multiply their science return. The SPB platform carries a $\sim$1200 kg science instrumentation payload at an altitude of 33 km for 70--100 days. At the time of this publication, two SPB missions have successfully flown from Wanaka, New Zealand. Compared to satellites, sub-orbital missions are inexpensive and benefit from late technology freeze: allowing their science payloads to incorporate the most recent developments in e.g.\ detector technology and lightweight mirrors. The science return of multiple balloon flights can rival or exceed that of high-profile UV-visible space missions at around 1\% of their cost. 

High-resolution observations from a stable platform are especially important to control systematic errors in cosmological probes based on weak gravitational lensing (WL): e.g.\ cosmic shear, galaxy-galaxy lensing, cluster/filament weak lensing, void lensing. WL methods rely on accurately measuring the shapes and redshifts of large ensembles of galaxies, most of which are small and faint. Controlling uncertainties in shape and redshift measurement is difficult from the ground, due to atmospheric distortions and limitations in wavelength coverage, in particular towards the blue/UV. Even a joint Euclid, LSST, and WFIRST data set would strongly benefit from SPB imaging. First, none of these missions offers the capability for space-quality resolution blue/UV imaging. Second, Euclid's wide-field survey is too shallow to calibrate shape measurement for LSST's faintest galaxies, and WFIRST's measurements are limited to the IR, where intrinsic morphologies are different. A deep, large-area SPB survey could fill the mismatches betwen these surveys, substantially reducing systematics. A deep, targeted SPB survey could follow up regions of interest, such as merging `bullet' clusters.

\begin{figure}
	\centering
	\includegraphics[width=\textwidth]{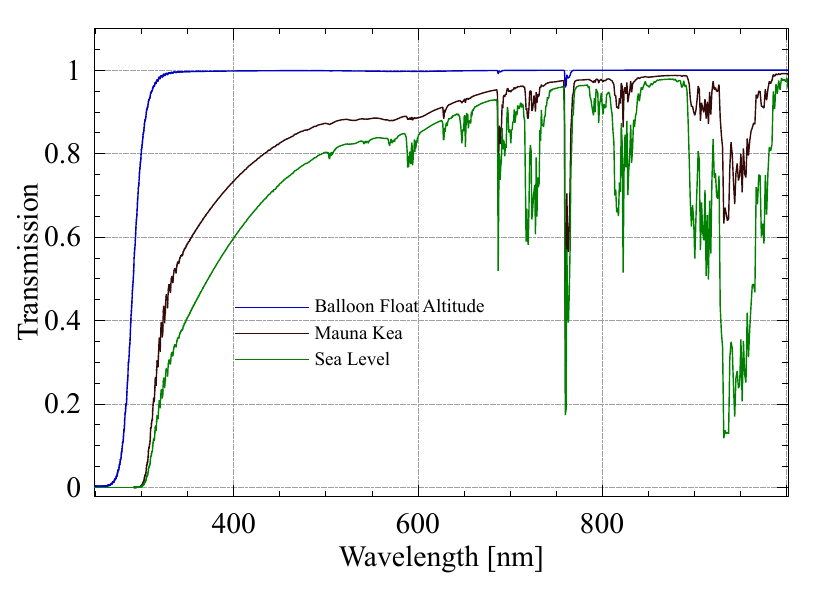}
	\caption{Atmospheric transmission as calculated by the MODTRAN4 software. It shows significantly reduced atmospheric absorption for a sub-orbital platform compared to even the best telescope sites on the ground. This advantage is particularly prominent for wavelengths below 400 nm.}
	\label{fig:balloontrans}
	\end{figure}
    
\subsection{High Precision Balloon-borne Astronomy}
At an altitude of 33-40 km, the sub-orbital balloon environment is ideal for astronomy in the optical and near-UV bands due to significantly reduced atmospheric interference (see Fig. \ref{fig:balloontrans}) when compared with ground-based systems. Additionally, the dramatically reduced cost of developing and launching balloon-borne systems when compared to similar space-borne systems is an attractive incentive for astronomers and cosmologists pursuing high-resolution imaging without exorbitant cost. As a result, mass and power restrictions for stratospheric instruments are often far less stringent than typical space systems (balloon-borne payloads on the order of 2-3 metric tons are feasible) with the added benefits of recoverability and reuse-ability of the balloon-borne payload at the end-of-life. 

Despite these clear advantages, balloon-borne instruments experience inherently unique challenges, which are attributed primarily to the nature of the balloon-borne environment and the launch vehicle, that are not present in typical ground or space-borne systems. Structurally, a typical stratospheric launch vehicle consists of a 1 million cubic metre helium balloon tethered to a 80-100 m long \textit{flight train} containing the parachute and attached to the scientific payload or \textit{gondola} through an actuated pivot connection (see Fig. \ref{fig:BIT_profile}). Subsequently, the dynamic environment is quite harsh, where compound pendulative effects dominate the low frequency regime with $\sim 0.05$ Hz pendulations about the balloon connection and $\sim 1$ Hz pendulations about the pivot connection. In addition to this, torsional effects through the flight train due to a slowly varying balloon rotation and stratospheric winds induce intermittent disturbances on the gondola throughout the flight. Thermally, an ambient temperature of roughly $-50^\circ$C to $-40^\circ$C gives rise to potential difficulties with maintaining alignment of sensitive optical components, where thermal gradients across the gondola can be detrimental to the overall optical performance and stability.  

In the field of balloon-borne optical astronomy, the most relevant past experiment is Stratoscope II, which was a telescope that demonstrated high pointing precision and image stability over several flights from 1967 -- 1973 \cite{RefWorks:95,RefWorks:88}. Overall, Stratoscope II achieved a 0.02$^{\prime\prime}$ focal plane equivalent pointing and image stability with an image resolution of 0.2$^{\prime\prime}$ for exposures from 5 to 20 seconds using a 36 inch telescope with an effective focal ratio of $f/105.2$ at the science camera \cite{Stratoscope_uranus}. At the time, this level of image stability was unmatched by any other optical instrument \cite{RefWorks:95, RefWorks:88} and effectively set the precedent for high precision astronomical instruments from the stratosphere. Despite its success, the Stratoscope II control architecture relied primarily on a low-gain system which compensated for stratospheric disturbances passively with occasional corrections on torque controlled axes \cite{RefWorks:88}. As a result, significant image processing post-flight was required to produce scientifically useful results at the quoted image and pointing stability, which severely limited the capabilities of the instrument. Therefore, there is precedent for the development of high-gain pointing and image stabilization systems for balloon-borne astronomy that provides on-demand and continuous correction of stratospheric disturbances for use in modern cosmological and astronomical experiments.

\section{BALLOON-BORNE IMAGING TELESCOPE (SUPERBIT)}
\label{sec:superbit}
The Super-pressure Balloon-borne Imaging Telescope, also known as \superbit, is an astronomical instrument designed to demonstrate the capabilities and performance of high-gain stability, on-demand, wide-field, visible-to-near-UV imaging from the stratosphere as a cost-effective, robust, and reusable platform for strong/weak lensing experiments as well as exo-planetary studies. As such, \superbit aims to provide a viable alternative to similar ground-based or space-borne systems, where the former suffers from atmospheric interference and the latter is often prohibitive in cost and development timescale. From an instrumentation perspective, \superbit is a 0.5 m Ritchey Cretien telescope with refractive field correction optics providing a $0.5^\circ$ field-of-view. The gondola, which houses, points, and stabilizes the telescope corrects for dynamic and pendulative disturbances with a three-axis gimbal system, which can provide telescope target acquisition to within < $1^\prime$ and telescope stabilization to within < $1^{\prime\prime}$ in all three axes. With additional correction from high bandwidth piezo-actuated backend optics, \superbit is capable of providing $0.2^{\prime\prime}$ resolution with $0.02^{\prime\prime}$ image stability over the full $0.5^\circ$ field of view with field rotation providing a maximum single exposure of over five minutes. 

\subsection{Mechanical Architecture}
\begin{figure}
	\centering
	\includegraphics[width=\textwidth]{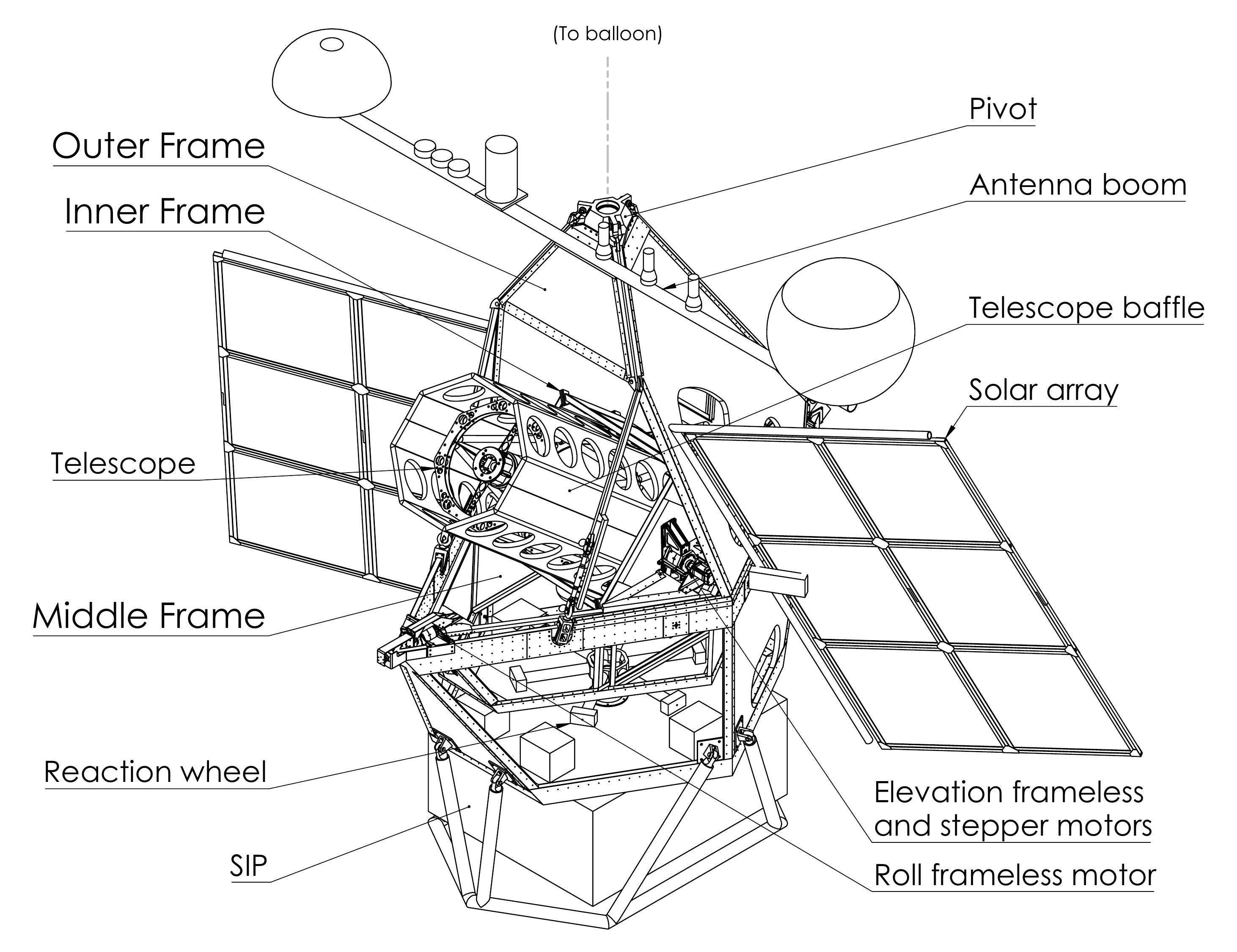}
	\caption{The BIT gondola is approximately 3 m tall from the base to the pivot joint and weighs approximately 1000 kg; the main structure is comprised of three independently rotating gimbals, constructed primarily out of aluminum honeycomb panels; during flight, the gondola communicates with the ground through the support instrumentation package (SIP), which is a modular isolated flight system given by the launch provider}
	\label{fig:BIT_profile}
	\end{figure}
The mechanical component of {\sc SuperBIT} is a mechanism that accommodates the pointing degrees of freedom, maintains the pointing accuracy, and withstands the various suborbital forces imposed on a balloon-borne telescope. Constructed primarily out of aluminum honeycomb panels, the {\sc SuperBIT} gondola (Figure \ref{fig:BIT_profile}) is approximately 3 m tall from the base to the pivot joint and weighs approximately 1000 kg.

	The telescope gondola has three free orthogonal axes to allow full rotational degrees of freedom. The mechanical design allows for the following ranges of pointing on the three axes -- pitch, roll, and yaw. In the pitch direction, the telescope points from $20^\circ$ to $55^\circ$ from the horizon. By design, the lower bound is limited by the Earth\rq s limb and the upper bound is determined by the angular extent of the full expansion of the helium balloon at cruise altitude. For the roll degree of freedom, the telescope must be able to rotate its view by around $\pm6^\circ$ to accommodate sky tracking over the duration of integration on the order of a few tens of minutes. Finally, the telescope is not be limited in the yaw direction as the gondola is free to rotate about the flight train axis.
	
	In order to achieve these degrees of freedom, three gimbal frames are used to allow telescope rotation in the three independent axes -- the Inner Frame, the Middle Frame, and the Outer Frame as shown in Fig. \ref{fig:BIT_profile}. Since the balloon is symmetric about the vertical axis, the yaw angle degree of freedom will be achieved by a pivot connecting the Outer Frame to the balloon tether. Inside the Outer Frame, the Middle Frame, is connected at the bow and stern of the Outer Frame to allow for the roll motion. And lastly, inside the Middle Frame, the Inner Frame, which rigidly holds the telescope, is connected at the port and starboard corners of the Middle Frame which allows for pitch movement. This equatorial mounting scheme accommodates all controlled degrees of freedom.

\subsection{Image Stabilization Hardware}
In order to stabilize the image on the telescope focal plane to within 0.02$^{\prime\prime}$, \superbit makes use of specific actuation hardware to control the gimballed axes with feedback from a number of inertial and sky-fixed pointing sensors. Architecturally, target acquisition can be broken down into three increasingly refined stages: coarse target acquisition, which includes large slews to a specified RA/Dec target on the sky to within $\sim1^\prime$; fine telescope stabilization, which stabilizes the telescope and tracks the specified target to within 1$^{\prime\prime}$; and back-end image stabilization, which further stabilizes the telescope focal plane to 0.02$^{\prime\prime}$. 

For coarse target acquisition, a large 16 kg$\cdot$m$^2$ reaction wheel controls gondola yaw, which roughly corresponds to telescope Az with respect to the horizon, with feedback from three fibre-optic rate gyroscopes on the telescope Inner Frame and a 3-axis magnetometer on the outer frame. In order to prevent reaction wheel saturation, a two-phase high resolution stepper motor spins the pivot at the top of the Outer Frame in order to dump momentum through the flight train to the balloon. On the gondola roll axis, a 16-bit absolute optical encoder is used to servo the Middle Frame to a desired offset roll initialization angle via two torque-controlled frameless DC motors and with rate gyroscope feedback. Lastly, the pitch axis is controlled via a dual-stage actuation scheme, where two coarse stepper motors servo the telescope frame to a desired pitch angle (corresponding roughly to telescope EL with respect to the horizon) while two additional frameless DC motors servo a pitch encoder with rate gyroscope feedback to a desired offset pitch initialization angle. Overall, the three actuated gimbal frames obtain coarse acquisition with respect to the target to within $\sim 1^\prime$ given the actuator and sensor characteristics in Tables \ref{table:sensor_char} and \ref{table:actuator_char}.

\begin{table}
\caption{\superbit attitude sensor characteristics \cite{Romualdez16}}
\label{table:sensor_char}
\begin{tabular}{*{1}{l}*{3}{c}r}
	\toprule
	Sensor Description & Readout Frequency (Hz) & Resolution & Noise Figure	\\
	\midrule
	Fibre optic rate gyroscope & 1000\textsuperscript{a} & $4.768\cdot 10^{-4}$ deg/s & $2.2\cdot 10^{-4}$ deg/(s$\cdot\sqrt{\mbox{Hz}}$)\\
	Absolute optical encoder & 100 & $5.49\cdot 10^{-3}$ deg & -- \\
	3-axis magnetometer & 20 & $6.7\cdot 10^{-5}$ Gs & $2.0\cdot 10^{-4}$ Gs\\
	Coarse elevation stepper & 10 & $9.374\cdot 10^{-3}$ deg & --\\
	Bore star camera & 20 & $0.23^{\prime\prime}$ centroids & $5.75\cdot {10^{-4}}^{\prime\prime}$/s\textsuperscript{b}\\
	Roll star camera & 20 & $0.46^{\prime\prime}$ centroids & $5.75\cdot {10^{-4}}^{\prime\prime}$/s\textsuperscript{b}\\
	Focal plane star camera & 60 & 0.025$^{\prime\prime}$ & $5.77\cdot {10^{-5}}^{\prime\prime}/\mbox{s}$\textsuperscript{b}\\
	\bottomrule
\end{tabular}
\\
\small{\textsuperscript{a} Asynchronous serial ($\pm 5\%$) remapped to synchronous 1000 Hz via Akima interpolation \cite{RefWorks:92}}
\\
\small{\textsuperscript{b} Sky equivalent read noise}
\end{table}
\begin{table}
\caption{SuperBIT actuator characteristics \cite{Romualdez16}}
\label{table:actuator_char}
\begin{tabular}{*{1}{l}*{2}{c}r}
	\toprule
	Actuator Description & Control Input & Characteristics\\
	\midrule
	Reaction wheel - frameless motor & 16-bit analog & 15 N$\cdot$m max.; 3600 lines/rev incremental encoder\\
	Pitch/roll - frameless motor $\times$ 4& 8-bit PWM\textsuperscript{a} & 5.0 N$\cdot$m max. torque; 3-phase Hall sensor feedback\\
	Pivot - stepper motor & step/direction & 0.018 deg/step\textsuperscript{b}; 2-256 \si\micro step/step; 0.44 N$\cdot$m hold \\
	Pitch - stepper motor $\times$ 2 & step/direction & 0.15 deg/step\textsuperscript{c}; 16 \si\micro step/step; 0.51 N$\cdot$m hold\\
	Piezo-electric tip/tilt & 16-bit analog & $0.027^{\prime\prime}$ resolution per axis; 45 Hz bandwidth\\
	\bottomrule
\end{tabular}
\\
\small{\textsuperscript{a} Pulse-Width Modulation}\\
\small{\textsuperscript{b} 1.8 deg/step motor through a 100:1 gear reducer}\\
\small{\textsuperscript{c} 1.8 deg/step motor through a 12:1 gear reducer}
\end{table}
From this coarse state, fine telescope stabilization with respect to the target on the sky is obtained primarily from two orthogonal star cameras. These star cameras capture images of single stars at 20 Hz, and relay their pixel centroid locations to the motor control system for feedback. As shown in Table \ref{table:actuator_char}, the star cameras can centroid to sub-arcsecond accuracies, and are sensitive to stars of at least Mag 9, which gives on average 4 stars per field of view. \cite{Zotov}. In addition to providing real time calibration of the gyro biases, the star cameras can also function as an absolute pointing sensor by taking degree-scale images and processing them using the ``astrometry.net'' suite of tools.\cite{dustin} This algorithm matches the observed star field against a database of 4-star distance patterns, and can provide an accurate position estimate in approximately two seconds on the flight computers. The feedback from both star cameras is used to stabilize the telescope via the frameless motors on the roll and pitch axes as well as the reaction wheel and pivot motor on the yaw axis. Note that coarse elevation stepper motors are not active within this level of stabilization because the granularity of the motor steps produce disturbances well above the 1$^{\prime\prime}$ level.

Once stabilized to 1$^{\prime\prime}$, the telescope back-end optics reduce the focal plane disturbances further to 0.02$^{\prime\prime}$ using a piezo-actuated tip/tilt fold mirror. High rate feedback for the tip/tilt actuator is provided by a focal plane star camera via a pick-off mirror from the primary telescope focal plane. This off-axis guider allows 60 Hz position feedback with minimal latency to the fine guidance system (FGS), and has 0.025$^{\prime\prime}$ resolution. With a diffraction limited spot size of around one pixel, the stars are oversampled and then filtered in software to increase the available centroiding precision. To compensate for potential thermal misalignment, the guide camera is affixed to a linear actuator which allows its focus to be adjusted independently of the science focal plane. 

\subsection{Control Architecture}

\begin{figure}
\centering
\includegraphics[width=0.8\textwidth,page=1]{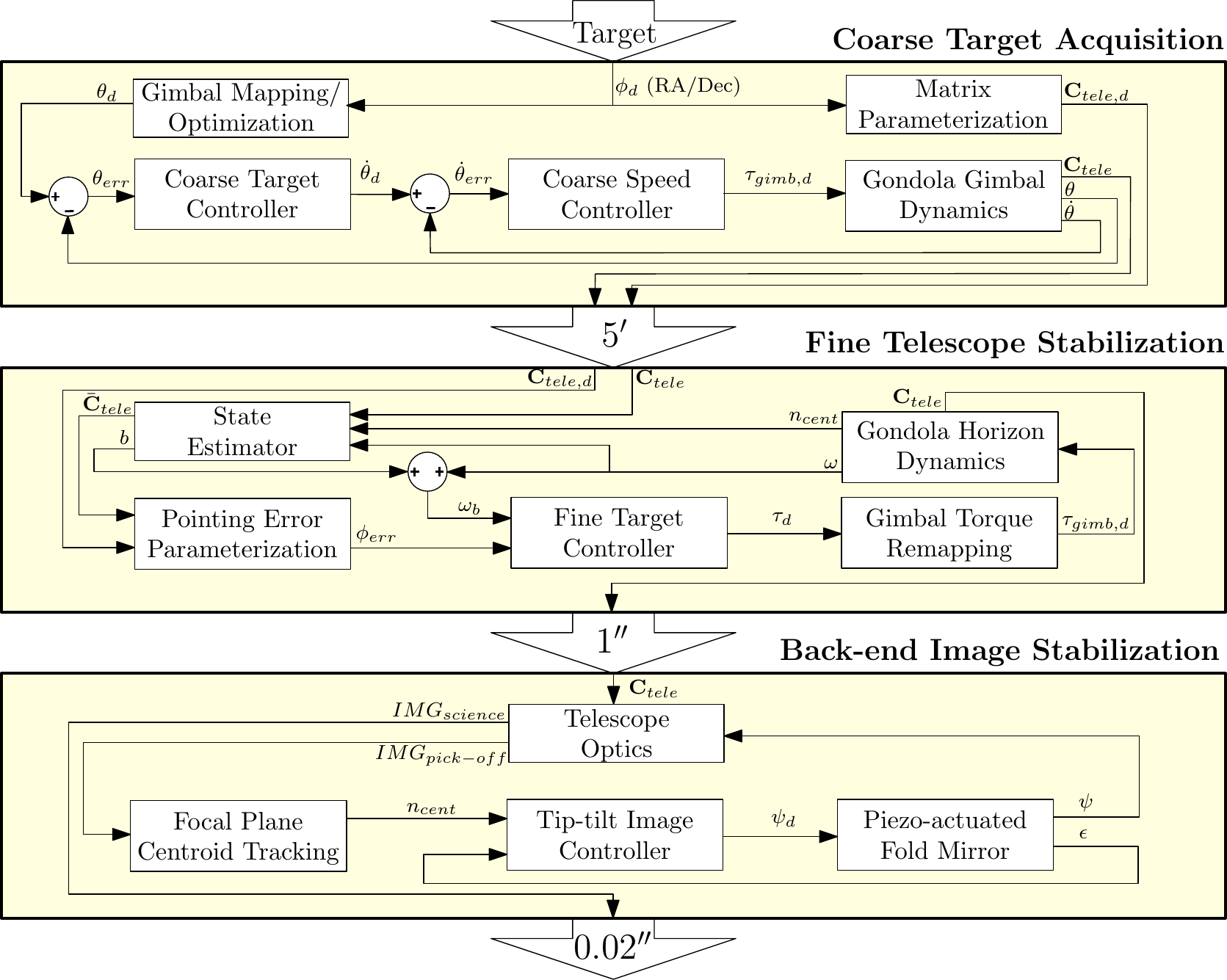}
\caption{SuperBIT high level control architecture for 0.02$^{\prime\prime}$ image stabilization on the telescope focal plane}
\label{fig:control_arch}
\end{figure}

Parallel to the pointing hardware, the \superbit control architecture can be broken down into three increasingly refined stages with a resulting image stability on the science focal plane of approximately 0.02$^{\prime\prime}$. A simplified block diagram of the control architecture is shown in Fig. \ref{fig:control_arch}. 

\begin{figure}
\centering
\includegraphics[width=0.5\textwidth,page=1]{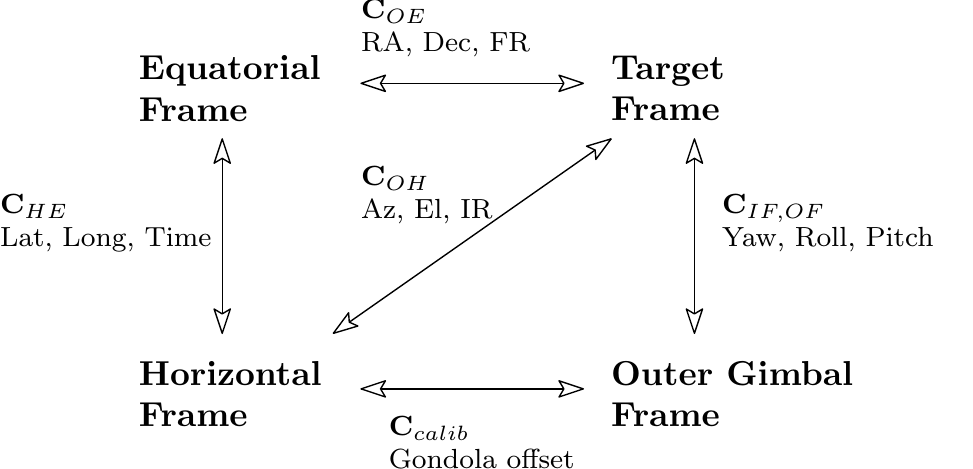}
\caption{Commutative diagram for relative coordinate transformations between the target frame $\coord{O} = \coord{IF}$, the equatorial frame $\coord{E}$, local horizontal frame $\coord{H}$, and the outer gimbal frame $\coord{OF}$; coordinate transformations are rotation matrices denoted by $\mathbf{C}$ which can be uniquely defined by their corresponding coordinates}
\label{fig:control_comm}
\end{figure}

Firstly, to facilitate smooth motion over large slews to a given astronomical target, the coarse target acquisition control scheme is comprised of a coarse target controller that issues a trapezoidal speed profile to the gondola gimbal control such that maximum gimbal acceleration and velocity can be set. For a given RA and Dec on the sky, the coordinates are projected onto the orthogonal gimbal axes for a given local latitude, longitude, and sidereal time through a series of calibration matrices to account for gondola imbalances and misalignments \cite{Romualdez16} (see Figure \ref{fig:control_comm}). Since the gimbal system is capable of full three-axis target acquisition with only two coordinates specified (i.e.\ RA/Dec), a nominal gimbal roll angle is chosen to maximize exposure time for the specified astronomical target, which roughly correlates with a fixed image rotation (IR) or field rotation (FR) on the sky \cite{Romualdez16}. Therefore, exposures are limited by the $\pm 6^\circ$ of the gimbal roll axis before having to reset with a different fixed field rotation for a particular target on the sky. 

Once sufficient coarse target lock has been achieved (< 1$^\prime$), coarse target acquisition control is deactivated to prevent stepper motors from causing disturbances, and the fine telescope stabilization control becomes active. Initially, target lock on the the bore sight star camera is achieved via the lost-in-space routine to determine the target acquisition error from coarse control. From this, high rate centroiding information from both roll and bore star cameras about the brightest stars in their respective fields-of-view is used to correct for coarse errors, stabilize the telescope, and track the the target to within 1$^{\prime\prime}$ ($1\sigma$). In order to ensure a high fidelity pointing solution, state estimation schemes use sensor data to estimate drifting rate gyroscope biases and to optimally combine centroiding information from both star cameras. Furthermore, the fine target controller implements globally valid gimbal torque mapping to ensure asymptotic stability during telescope stabilization and is necessary for robust control. 

At this level of stability, the brightest tracking star on the focal plane star camera is used as feedback for the piezo-actuated tip-tilt fold mirror to reduce remaining disturbances up to 45 Hz down to 0.02$^{\prime\prime}$ on the science focal plane. Subsequently, full target lock is achieved and exposures for the specified target can begin. Since both telescope and image stabilization are actively and concurrently stabilizing the focal plane, leakage of high frequency resonant modes from the fine telescope control can be mitigated through selective filtering in focal plane centroid feedback, so that structural resonances in the gondola structure do not degrade image stability.

\section{ENGINEERING TEST FLIGHTS}
\label{sec:testflights}

\subsection{Timmins 2015 Flight}

The inaugural flight of the Balloon-borne Imaging Testbed (aka.\ BIT) took place from September 18--19, 2015. It was  launched from the Timmins Stratospheric Balloon Base in Ontario, Canada, by the Canadian Space Agency (CSA) and the Centre National d'\'Etudes Spatiales (CNES). Communications to and from the BIT gondola used a high-gain line-of-sight link, 
allowing for human-in-the-loop teleoperation throughout the flight. BIT remained at 36 km altitude for 6.5 hours, and was safely recovered with minimal damage, following descent by parachute.

The primary goal of this flight was to test the engineering capabilities of BIT's pointing system: in particular the telescope stabilization and pointing precision, as well as fine image stabilization of the back-end optics. It was thus designed as a proof-of-concept for \superbit, as well as a way to characterize system performance within the harsh dynamic and thermal environment of the upper atmosphere.

\begin{figure}
\centering
\includegraphics[width=0.8\textwidth,page=1]{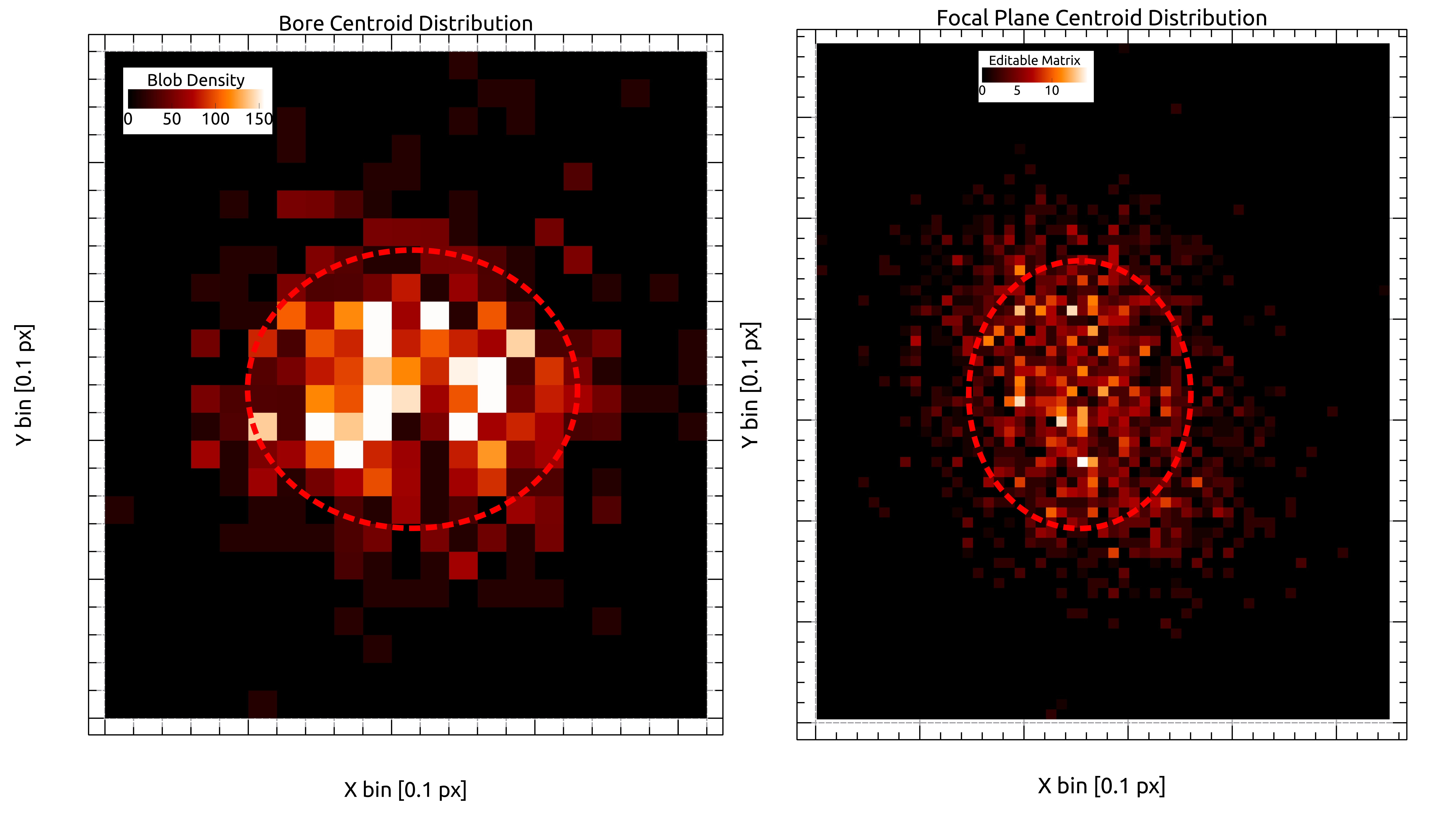}
\caption{Fine pointing stabilization during the BIT 2015 Timmins flight as demonstrated by the centroid dispersion in the bore (left) and focal plane (right) star cameras over a 23 minute tracking period\cite{Romualdez16}; the bore and focal plane star cameras have a sky-equivalent scale of approximately 0.23$^{\prime\prime}$ and 0.025$^{\prime\prime}$ per tenth of a pixel, respectively; the $3\sigma$ ellipses are shown in red dashed lines, where the $1\sigma$ values for the pointing stability in the bore and focal plane cameras are 0.68$^{\prime\prime}$ and 0.12$^{\prime\prime}$, respectively}
\label{fig:2015_pointing_stab}
\end{figure}

The flight schedule included several tracking runs, during which coarse target acquisition, fine telescope stabilization, and image stabilization were tested. The telescope was successfully stabilized to within $0.68^{\prime\prime}$ ($1\sigma$) over a number of 1.4 hour tracking runs, as verified by the centroid dispersion in the bore star camera in Fig.~\ref{fig:2015_pointing_stab}. From an attitude determination perspective, state estimator consistency agreed well with estimated pointing errors, where pointing stability was limited mostly by the rate of gyroscope integrations between star camera measurements at 3 Hz. 

During several of these tracking runs, a star on the focal plane star camera was used to obtain fast feedback for additional back-end optical stabilization. During stabilization, it was discovered that telescope optics were misaligned, resulting in an asymmetric point spread function, even at optimal focus. Furthermore, it was observed that the resulting stabilized telescope environment had unexpected residual disturbances beyond the bandwidth of the piezo-electric tip-tilt fold mirror, which reduced the desired performance. Despite these shortcomings, however, back-end optical stabilization further stabilized the focal plane down to $0.12^{\prime\prime}$ ($1\sigma$) as shown in Fig. \ref{fig:2015_pointing_stab}. 

Overall, the 2015 Timmins flight successfully demonstrated many of BIT's engineering capabilities: esepecially the on-demand, continuous, and robust telescope pointing stability. 
The limitations in achieved optical performance have informed design decisions and changes required for \superbit. These include the need for a higher bandwidth piezo-actuator, plus 
in-flight calibration and actuation of optical elements.

\subsection{Palestine 2016 Flight}

To improve on the optical performance from the 2015 Timmins flight, \superbit now includes upgrades to both hardware design and in-flight procedures. To start, a higher bandwidth piezo-electric tip-tilt actuator (> 40 Hz) allows for a reduction of the high frequency dynamic disturbances observed in the previous flight, while allowing for more margin in the controller parameters for fine telescope stabilization. In effect, higher bandwidth correction on the focal plane allows for a greater overlap in control with the coupled fine telescope stabilization hardware such that trade-offs can be made between the two subsystems when optimizing controller gains. 

A more rigorous procedure for optical alignment was also developed for the pre-flight integration. Specifically, the telescope's primary and secondary mirrors were aligned using an external monochromatic point source and parabolic collimator mirror, the shape and size of the point-spread function was characterized throughout the focal plane, and telescope collimation was verified under various operational conditions (i.e.\ different elevations, loading, etc.). To maintain alignment in the stratospheric environment, where thermal gradients to a $-40^\circ$C ambient temperature could degrade optical performance, a temperature controlled shroud fully encloses the telescope assembly. At the time of publication, \superbit\ is waiting for suitable weather for launch.

\section{OPERATIONAL FLIGHTS}
\label{sec:operationflights}

\subsection{Super Pressure Balloon (SPB) Flight}

\superbit is scheduled for a 100 day super presure balloon flight from Wanaka, New Zealand by 2018. 
From an engineering perspective, the improvements that are essential to maintaining \superbit's optimal performance over three months centre on increased autonomy and robustness of existing systems. 
The \superbit gondola must self-regulate power and thermal subsystems over day-night cycles, while preventing damage to sensitive optical components during inactivity in the day.
Electronic subsystems (flight computers and memory storage elements) must be robust to cosmic rays and other errors associated with the increased radiation environment in the stratosphere. 
Telescope optics must maintain collimation, through autonomous aquisition and observation of bright stars or on-board metrology, followed by 6 degree of freedom actuation of the secondary mirror and a $z$-stage actuator for the focal plane star camera.
Lastly, the ability and redundancy to downlink science flight data is imperative to reduce the risk of data loss either during flight or in the event that the payload is irrecoverable. 

The primary scientific focus of the \superbit SPB flight will be to obtain high-precision weak and strong gravitational lensing measurements of $\sim$200 galaxy clusters at redshifts $0.1<z<0.5$, and their attachment to the cosmic web.
\superbit's cluster sample is intentionally compiled from a variety of X-ray, Sunyaev-Zel'dovich and optical/NIR imaging surveys.
Cross-calibration of cluster mass measurements by a single, high-precision instrument will resolve current discrepancies between cosmological tests (which may stem from foregrounds in the CMB, mis-calibration of X-ray telescopes, or previously unknown physics).
The increasing number and mass of clusters towards low redshift (reflecting the gradual formation of cosmic structure) is a sensitive probe of the nature of Dark Energy.
The trajectory of Dark Matter through collisions between galaxy clusters reveals the interaction properties of Dark Matter. The steady-state distrubtion of Dark Matter long after a collision reveals its interaction properties and dynamical temperature. At any time, roughly half of clusters are observed during collisions, and half are observed in a relaxed state.

\subsection{Facility Class Instrument}
\superbit is a step toward a future platform that can accommodate facility-class instruments requiring sub-arcsecond stability for annual suborbital flights of up to 100 days in the early 2020's. Such a mission will be able to carry lightweight mirrors between 1 and 2 meters.
In combination with a giga-pixel class focal plane (compared to 32MP for \superbit), these systems will exceed the imaging capability of HST and will play a highly flexible role among the fixed concert of Euclid, WFIRST and LSST.
Below we identify some of the broad advantages of stratospheric balloon-based observations that motivate long-term development of a flexible observing platform:
\paragraph*{UV photometry:} Strong UV absorption by the atmosphere makes wide-field near-UV imaging from the ground restrictively inefficient and time consuming. However, these data are critical for the accurate determination of photometric redshifts that are needed, for instance, in dark energy studies. A wide-field UV imaging survey would therefore benefit both Euclid and WFIRST. As illustrated in Fig.~\ref{fig:balloontrans}, observations from stratospheric altitudes suffer significantly less UV absorption and atmospheric background than from the ground. A meter-class telescope, even coupled to a modest camera, has similar survey speed to LSST in its bluest band, but at a much higher resolution (and extending to 300\,nm wavelength) -- allowing the terrestrial telescope to focus on the visible and near-IR.
\paragraph*{Persistent, sub-arcsecond imaging:} 
As demonstrated by the ongoing, $8\times$ oversubscription of HST, even 25 years after its launch, many branches of astronomy exhibit a persistent need for high resolution UV/optical/IR observations. For example, observations of low-surface-brightness galaxies (and satellites of our own Milky Way) and a census of the stars within the Galaxy and nearby galaxies (galactic archaeology) would all greatly benefit from the high resolution of an optical-UV balloon-borne telescope. Exoplanet searches, such as microlensing studies of stars toward the Galactic bulge, also benefit from high-resolution imaging, over a long time baseline of observations. A balloon-borne microlensing program covering, e.g. some of the proposed WFIRST microlensing fields would not only provide added science value to that mission by stretching this baseline, but also be immune to the weather-induced losses of observations that can interrupt the critical cadence needed for these observations. 
\paragraph{Technological development:} The path to developing new sensors, detectors, and instruments for space missions is difficult and expensive. Few opportunities exist to fly new technologies in a space-like environment at a reasonable cost. A facility class instrument flying regularly on the SPB platform with a relatively low launch cost, will provide a technology testbed to help ensure space qualification of components and instruments that will become the backbone of future flagship, probe, and Explorer-class space missions.

\section{CONCLUSIONS}
\label{sec:conclusions}
Progress in many branches of cosmology and astrophysics currently relies on space-quality observations. From below the Earth's turbulent atmosphere, astronomy is limited by the blurring or blending of adjacent sources, and the misidentification of Milky Way stars with distant galaxies. Recent advances in multi-conjugate adaptive optics have helped some of the science cases that require only a narrow field-of-view. However, rising above the atmosphere will remain the only solution for science cases that require accurate knowledge of the PSF, or a wide field of view. 

We report the successful demonstration of a balloon-borne telescope, BIT, flying above 99\% of the Earth's atmosphere. For less than 1\% the cost of an equivalent satellite, but in conjunction with advances in `superpressure' balloon technology that recently extended flight duration from $\sim3$ days to $\sim3$ months, this offers transformative opportunities. \superbit is scheduled for a ultra-long duration, 100 day flight by 2018.

This is merely a first step toward an ambitious but achievable goal of facility-class 1--2 m telescopes providing hundreds of days of near space-quality imaging and spectroscopy. These missions will be a cutting edge but flexible facility, able to rapidly adapt to the most interesting science areas throughout the 2020s. Of particular interest will be the interplay of these SPB missions with large ground and space-based missions, such as Euclid, LSST, and WFIRST. As well as quickly following up currently unimagined science goals, \superbit's UV and high resolution complementarity offers a unique potential to enhance science return even in key research areas including exoplanets, dark energy and dark matter science.

\acknowledgments
\superbit is supported in Canada, via the Natural Sciences and Engineering Research Council (NSERC), in the USA via NASA award NNX16AF65G, and in the UK via the Royal Society and Durham University's astronomy projects grant. Part of the research was carried out at the Jet Propulsion Laboratory (JPL), California Institute of Technology (Caltech), under a contract with NASA.

\bibliography{report} 
\bibliographystyle{spiebib} 

\end{document}